\documentstyle[12pt,a4wide,epsfig]{article}
\newcommand{\XLOOPS}{{\em XLOOPS}}
\newcommand{\BXLOOPS}{{\em\bf XLOOPS}}
\newcommand{\ssigma}{\hbox{$\kern2.5pt\vrule height4pt\kern-2.5pt\sigma$}}

\newcommand{\oone}{\hbox{$1\kern-2.5pt\hbox{\rm l}$}}

\newcommand{\slD}{\kern1pt/\kern-8pt D}
\newcommand{\slk}{/\kern-6pt k}
\newcommand{\sll}{/\kern-4pt l}
\newcommand{\slp}{p\kern-5pt/}
\newcommand{\slq}{q\kern-5.5pt/}
\newcommand{\sls}{s\kern-5pt/}
\newcommand{\slu}{u\kern-5.5pt\raise1pt\hbox{$\scriptstyle/$}\kern1.5pt}
\newcommand{\slv}{v\kern-5pt\raise1pt\hbox{$\scriptstyle/$}\kern1pt}
\newcommand{\Frac}[2]{\hbox{$\frac{#1}{#2}\ $}}

\begin{document}
\thispagestyle{empty}
\begin{flushright}
MZ-TH/97-32\\
hep-ph/9709490 \\
September 1997\\
\end{flushright}
\vspace{0.5cm}
\begin{center}
{\Large\bf Massive Two-Loop Integrals and}\\[.3cm]
{\Large\bf Higgs Physics}\\[1.3cm]
{\large A.~Frink, J.G.~K\"orner and J.B.~Tausk}\\[1cm]
Institut f\"ur Physik, Johannes-Gutenberg-Universit\"at,\\[.2cm]
Staudinger Weg 7, D-55099 Mainz, Germany\\
\end{center}
\vspace{1cm}

\begin{abstract}\noindent
We present an overview of the research activities of the theoretical
particle physics group at the University of Mainz on the calculation
of massive one- and two-loop Feynman diagrams. The main objective of
this research was to develop an automatic one- and two-loop calculation
program package.  A first version of such a program was recently realized
by Br\"ucher, Franzkowski and Kreimer.  We describe in some detail the
present features of this automatic loop calculation program as well as
the integration techniques that go into the program.  The program runs
under the name of \XLOOPS. The present version is labelled by \XLOOPS\
1.0. The program allows one to calculate massive one- and two-loop
integrals in the Standard Model including their tensor structure - all
at the click of a mouse. Renormalization features are included in that
the UV divergences in UV divergent integrals are explicitly computed in
dimensional regularization. One-loop integrals are calculated analytically
in $d\neq4$ dimensions whereas two-loop integrals are reduced to two-fold
integral representations which the program evaluates numerically. We
attempt to provide a synopsis of the novel loop integration techniques
that have been developed for the \XLOOPS\ program. They allow for a fast
and efficient evaluation of massive one- and two-loop integrals. We
discuss Higgs decay at the two-loop level as a first application of
the novel integration techniques that are incorporated into \XLOOPS\ .
\end{abstract}

\newpage\noindent

\section{Introduction}

In the last few years a large part of the theoretical particle physics
group at the University of Mainz has been involved in the computation
of massive one- and two-loop Feynman integrals. This activity was
part of a continuing group seminar series on the evaluation of Feynman
diagrams. Apart from the work of the senior members of the group many
beautiful new results have been achieved by our diploma and PhD students
in the context of their thesis work. Present members of the group
are L. Br\"ucher, M. Fischer, J. Franzkowski, A. Frink, S. Groote,
V. Kleinschmidt, J.G. K\"orner, R. Kreckel, D. Kreimer, A.J. Leyva,
M. Mauser, K. Schilcher and J.B. Tausk. Past members of the
group include A. Czarnecki, U. Kilian, K. Melnikov, D. Pirjol, P. Post and
O. Yakovlev. We have greatly benefitted from the many visitors and
guest scientists who have visited our group over the years and who
have generously shared with us their insights into the calculational
techniques of Feynman diagram computations.  They are too numerous to be
all quoted here in name. All of this effort would not have been possible
without the support of various funding agencies. These include the DFG
(Deutsche Forschungsgemeinschaft) with individual grants as well as
stipends provided by the Graduiertenkolleg ``Teilchenphysik bei hohen
und mittleren Energien", the BMBF (Bundesministerium f\"ur Bildung und
Forschung) and a HUCAM project granted by the European Union.

This write-up is certainly in the character of a workshop write-up.
We take this opportunity to present an overview of the activities of our
group and the results obtained therein. In particular this means that
we do not attempt to give a balanced account of the present state of
art of loop calculations the world over. We ask for forebearness. It
is clear that the results that we are presenting are too numerous to
have been obtained by any single person alone. The results represent the
efforts of various subsets of the members of the group. When discussing
particular results we shall not always specify the subset of members of
the group that were involved in a particular calculation in order to avoid
excessive cross-referencing. The respective members can be identified
from the references.

By means of introduction we want to remark that there are two main
approaches to doing loop calculations. The one approach we would like
to call the absorptive-dispersive approach. One first calculates the
absorptive parts of a Feynman diagram and then uses dispersion relations
with possible subtractions to obtain the dispersive part of the Feynman
diagram. The experience is that the absorptive-dispersive approach
usually goes further when calculating two-loop diagrams if the aim is to
perform analytically as many of the necessary integrations as possible.
This is particularly true for two-loop diagrams with internal loop masses.

\begin{figure}[htbp]
 \begin{center}
    \epsfig{file=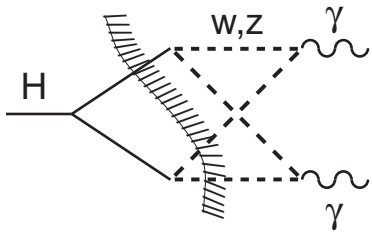,width=0.3\linewidth}
 \end{center}
 \caption[]{Three-particle cut of a crossed ladder diagram contributing
            to $H \to \gamma+ \gamma$. Solid lines: mass Higgs particles;
            dashed lines: massless gauge bosons of the effective theory.}
 \label{fig:hgg}
\end{figure}

There are many examples of two-loop calculations that have been done
using the absorptive-dispersive integration technique. For example,
in Mainz we have calculated the two-loop gauge boson contributions to
the decay $H\rightarrow \gamma+ \gamma$ using the absorptive-dispersive
technique \cite{KoernerMelnikovYakovlevHgg}. The calculations were done
using the Equivalence Theorem. In the high energy limit the transition
$H\rightarrow \gamma+ \gamma$ is dominated by the longitudinal degrees
of freedom of the internal loop gauge bosons which can be represented
by the massless and spinless Goldstone excitations of an effective
scalar theory. The Higgs boson is the only massive particle in the
theory. It appears both as external and internal loop particle. There
are a number of two-loop diagrams that contribute to the process. They
are too numerous to be all shown here. All absorptive contributions were
determined analytically, and all but one of the remaining dispersive
integrals were done analytically. In Fig.\ref{fig:hgg}
we show the contribution of the three-particle cut of the the massive
two-loop crossed ladder three-point function which was the only
contribution which did not yield to analytical integration in the
dispersion integral. The remaining one-dimensional dispersion integral
of this one particular contribution was then done numerically. One of
the outcomes of the analysis was that the results of the effective
theory should be reliable in the range $0.6\,\mbox{TeV} \leq m_H \leq
1.5\,\mbox{TeV}$. Above $1.5\,\mbox{TeV}$ the two-loop result blows up.

In another example the absorptive-dispersive approach was used to
calculate the two-loop contributions to the $K^0$ charge radius in chiral
perturbation theory \cite{Post}. Quite naturally the loop
masses in chiral perturbation theory cannot be neglected since chiral
perturbation theory is an effective low energy theory. The final result
of the calculation was obtained in terms of a one-dimensional integral
representation which was evaluated numerically.

\begin{figure}[htbp]
 \begin{center}
    \epsfig{file=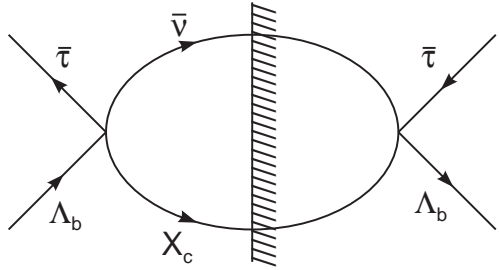,width=0.4\linewidth}
 \end{center}
 \caption[]{Rate of $\Lambda_b^{(\uparrow)}
                \to X_c+\tau^{-(\uparrow)}+\bar{\nu}_{\tau}$
            extracted from the absorptive part of the one-loop amplitude
           $\tau^{+(\uparrow)} + \Lambda_b^{(\uparrow)} \to
            \tau^{+(\uparrow)} + \Lambda_b^{(\uparrow)}$ .}
 \label{fig:hqet}
\end{figure}

As a curious side remark we want to emphasize that it is even sometimes
advantageous to take the opposite route and first calculate the
full loop contribution directly and then extract its imaginary part.
For example, rate calculations can be done quite efficiently in this
way.  If one shies away from phase-space integrations but knows how
to do loop calculations e.g. by Feynman parameter methods then one can
obtain rates very efficiently from the imaginary part of the full loop
contribution. This technique has been used to great advantage in the
calculation of the inclusive semileptonic differential rate of polarized
$\Lambda_b$'s to polarized $\tau$'s within the Operator Expansion method
in HQET \cite{BalkKoernerPirjol}. The relevant one-loop diagram is shown
in Fig.\ref{fig:hqet} together with its absorptive cut which determines
the rate of the decay.

A large part of the remaining discussion in this presentation will concern
itself with the direct computation of massive two-loop integrals not
using the absorptive-dispersive approach.  As a general feature of the
direct approach we shall find that all but two of the massive two-loop
integrations can be done analytically. In the absorptive-dispersive
approach there are many examples where one can go further and where all
but one of the massive two-loop calculations can be done analytically. A
noteworthy example is the efficient one-dimensional integral
representation written down by Bauberger and B\"ohm for the general
massive two-loop two-point function \cite{BaubergerBoehm}. This has to be
compared with the two-fold integral representation derived in the direct
approach to be discussed later on. However, the absorptive-dispersive
approach involves a detailed consideration of the two- and three-particle
cut structure of the contributing diagrams including possible subtraction
terms and anomalous threshold contributions which is not always simple
theoretically. In particular the requisite analysis does not possess
the flexibility which is needed for a general purpose program package
that we were aiming for.

\section{Direct Approach to Loop Calculations}

In Mainz we have developed a completely new approach to the calculation
of massive one- and two-loop diagrams. These radically new integration
techniques have been devised in a series of ingenious papers by Kreimer
\cite{Kreimer2L2P,Kreimer1L,KreimerOther2L,Kreimer2LTensors}. In order
to appreciate the quality of innovation of these new techniques let us
briefly recapitulate the main features of the more traditional approach
to loop integrations. The essential integration features employed by most
practitioners in the field may be summarized by the following statements

\begin{itemize}
\item integrate covariantly
\item introduce a set of Feynman parameters
\item perform a Wick rotation and integrate in Euclidean space
\item do tensor integrals \`a la Passarino-Veltman
\end{itemize}

In comparison to the above program the Mainz method may be characterized
by the following list of statements

\begin{itemize}
\item integrate in a special frame (usually the rest frame of one of the
external particles)
\item no Feynman parameters
\item do the integrations in Minkowski space
\item tensor integrals are worked out by direct integration
\end{itemize}

The traditional integration methods become progressively
more intractable when more and more mass scales are introduced
into a problem. This has led many authors to introduce quite radical
approximations such as setting the ratios $m_W/m_t$ and $m_Z/m_t$
to zero in their loop calculations without being able to say much about
the error invoked by such an approximation. Contrary to this
the Mainz integration methods are ideally suited for the most general
mass case with as many different massive loop particles as possible.
In fact Standard Model calculations with $m_t \neq m_W \neq m_Z \neq m_H$
are quite welcome.
In the other extreme one generally has no problems taking the mass zero
limit in a loop contribution if the diagram does not contain any infrared
(IR) or mass (M) singularities.
In the presence of IR/M singularities the corresponding
integrals have to be regularized by subtraction as will be discussed later
on.

\begin{figure}[htb]
  \begin{center}
    \epsfig{file=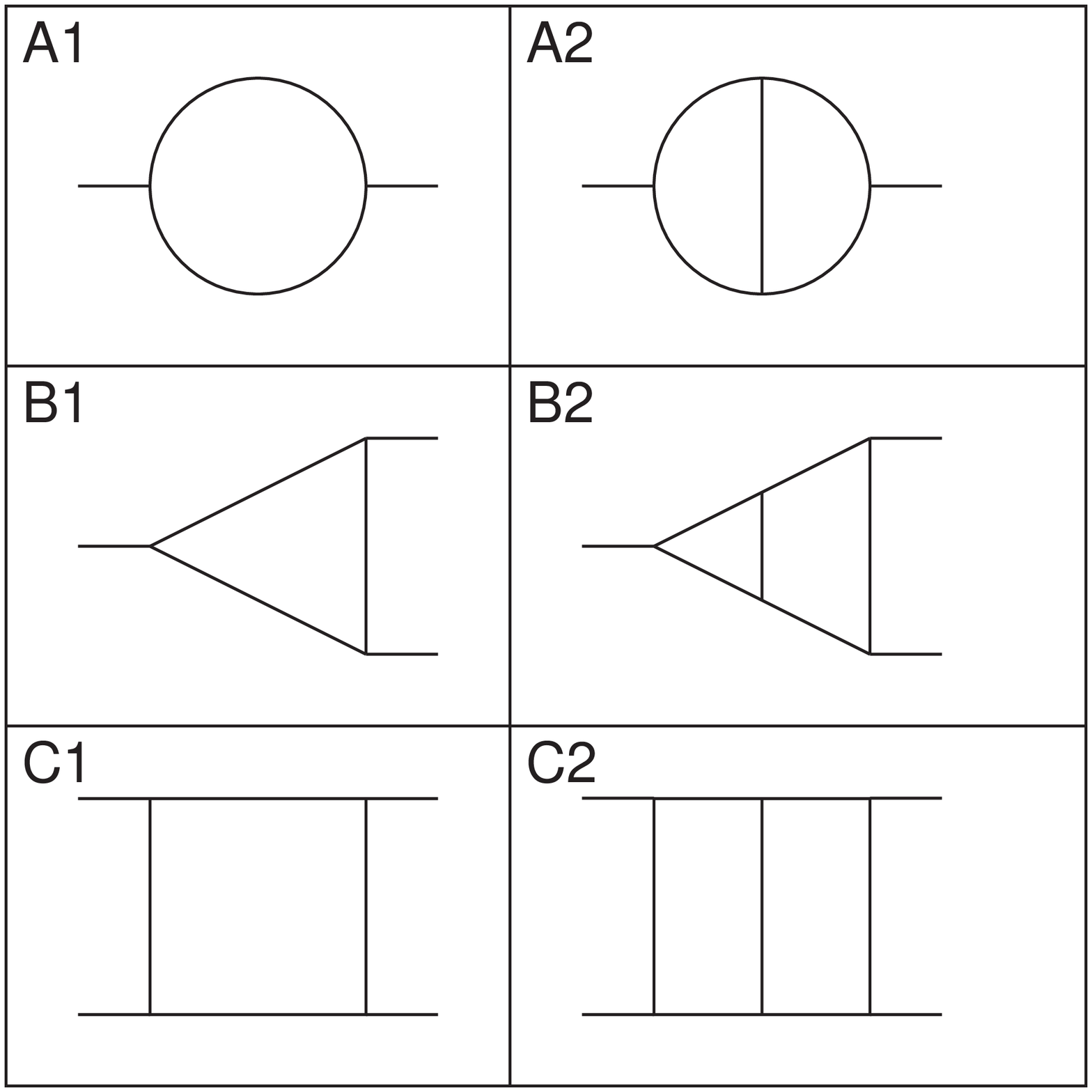,width=0.4\linewidth}
  \end{center}
 \caption[]{Representative set of diagrams to be treated in \XLOOPS}
 \label{fig:abc12}
\end{figure}

In order to provide a measure of the scope of what \XLOOPS\ \cite{XLoops}
can do at the moment and what \XLOOPS\ hopes to be able to do in
the future we have drawn a representative set of loop diagrams in
Fig.\ref{fig:abc12} which are labelled
as: A1 (two-point one-loop), A2 (two-point two-loop),
B1 (three-point one-loop), B2 (three-point two-loop),
C1 (four-point one-loop) and C2 (four-point two-loop).
In Table 1 we list the status of these loop calculations as concerns their
availability in the present version \XLOOPS\ 1.0 and, if not implemented
in \XLOOPS\ 1.0, we comment on their stage of development.

\begin{table}[htb]
\begin{center}
\begin{tabular}{|l|l|}\hline
loop class     & status and availability\\ \hline
2-point 1-loop & \XLOOPS\ 1.0\\ \hline
2-point 2-loop & \XLOOPS\ 1.0\\ \hline
3-point 1-loop & \XLOOPS\ 1.0\\ \hline
               & UV divergences    : ready\\
3-point 2-loop & IR/M divergences : starting\\
               & tensor structure  : conceptually ready\\ \hline
4-point 1-loop & ready for implementation\\ \hline
4-point 2-loop & starting\\ \hline
\end{tabular}
\end{center}
\caption[]{Present status of one- and two-loop calculations in \XLOOPS}
\end{table}

\section{Parallel and orthogonal momentum space}

A very important technical concept in our loop calculations is the
splitting of total $d$-dimensional space into a parallel space spanned
by the external momenta and an orthogonal space spanned by its
orthogonal complement. Symbolically we write
\begin{quote} total space = parallel($\parallel$) space
+ orthogonal($\perp$) space
\end{quote}
The orthogonal space components of the loop momenta have no orientation
in the real world defined by parallel space. For scalar integrals
this implies a spherical symmetry in orthogonal space which can be
nicely exploited. Tensor integrals involving orthogonal components
can easily be evaluated since their expansion along resultant tensors
involves only combinations of orthogonal space metric tensors. These
can be easily processed further. Also, by splitting $\gamma$-matrices
into their orthogonal and parallel space components almost all the
tedious $\gamma$-matrix algebra in fermionic amplitudes can be avoided.

\begin{figure}[htbp]
  \begin{center}
    \epsfig{file=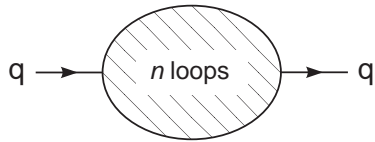,width=0.3\linewidth}
  \end{center}
 \caption[]{Two-point function with any number of loops}
 \label{fig:nloop2pt}
\end{figure}

Let us illustrate the advantages of working in orthogonal space
with the help of a few examples. We begin with the two-point function
with any number of loops as drawn in symbolic fashion in
Fig.\ref{fig:nloop2pt}.

The external momentum and thus parallel space is defined by the momentum
$q_\mu$. The dimensionality of the two respective spaces is thus given by
\begin{center}
\begin{tabular}{lcl}
parallel space $(\parallel)$ & : & 1-dimensional\\
orthogonal space $(\perp)$   & : & $(d-1)$-dimensional\\
\end{tabular}
\end{center}

The metric tensor $g_{\mu\nu}$ splits up into a parallel piece
$g_{\mu\nu}^\parallel$ and a orthogonal piece
$g_{\mu\nu}^\perp$ according to $g_{\mu\nu}=g_{\mu\nu}^\parallel
+g_{\mu\nu}^\perp$. The respective components of the metric tensor
can easily be constructed, $viz$.
\begin{eqnarray}
g_{\mu\nu}^\parallel & = & q_\mu q_\nu/q^2\\
g_{\mu\nu}^\perp & = & g_{\mu\nu}-q_\mu q_\nu/q^2
\end{eqnarray}
Quite naturally the two components of the metric tensor can be used
to project out the parallel and orthogonal components of any $d$-dimensional
vector.

In a similar way one can split up the $\gamma$-matrix into its
parallel and orthogonal components according to
\begin{equation}
\gamma_\mu=\gamma_\mu^\parallel+\gamma_\mu^\perp
\end{equation}
By using the metric projectors it is easy to see that one has
$\gamma_\mu^\parallel=\slq q_\mu/q^2$ and
$\gamma_\mu^\perp=\gamma_\mu - \slq q_\mu/q^2$. Note also that
$\gamma_\mu^\parallel$ and $\gamma_\mu^\perp$ anticommute, i.e.
one has
\begin{equation}
\left[ \gamma_\mu^\parallel,\gamma_\nu^\perp \right]_+ = 0
\end{equation}
The anticommutativity of the two respective $\gamma$-matrix components
results in an extremely expedient way of dealing with the $\gamma$-matrix
algebra when evaluating fermionic Feynman diagrams. We shall return
to this point later on.

\begin{figure}[htbp]
  \begin{center}
    \epsfig{file=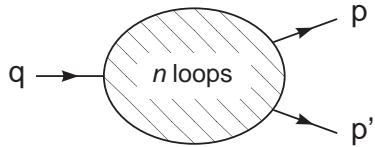,width=0.3\linewidth}
  \end{center}
 \caption[]{Three-point function with any number of loops}
 \label{fig:nloop3pt}
\end{figure}

In the case of the $n$-loop three-point function one has two independent
external momenta $q$ and $p$ that define parallel space. A symbolic
representation of the $n$-loop three-point function is drawn in
Fig.\ref{fig:nloop3pt}.

Assuming that the momentum $q$ is time-like one chooses the rest frame
of $q$ as the reference frame (if $q$ is space-like one chooses the rest
frame of one of the other momenta).
In this frame the momenta $q$ and $p$ are represented
by $q=(q^0,0,0,0)$ and $p=(p_0,p_z,0,0)$ such that the three-momentum
$\vec{p}$ defines the $z$-direction in parallel space.
The dimensions of the two respective spaces is now given by
\begin{center}
\begin{tabular}{lcl}
parallel space $(\parallel)$ & : & 2-dimensional\\
orthogonal space $(\perp)$   & : & $(d-2)$-dimensional\\
\end{tabular}
\end{center}

As before the metric tensor splits into the the two components
\begin{equation}
g_{\mu\nu}=g_{\mu\nu}^\parallel + g_{\mu\nu}^\perp
\end{equation}
where one has
\begin{equation}
g_{\mu\nu}^\parallel=\frac{1}{q^2} q_\mu q_\nu -
\frac{1}{p_z^2}\left(p_\mu p_\nu-\frac{(pq)^2}{q^4}q_\mu q_\nu\right)
\end{equation}
and $g_{\mu\nu}^\perp$ is given by the complement
$g_{\mu\nu}-g_{\mu\nu}^\parallel$.

We briefly side-track to remind the reader that the construction of
a parallel and an orthogonal space is a familiar concept used also in
other areas of particle physics even if different names are attached to
the two respective spaces. For example, one talks of the longitudinal
and transverse component of a four-vector current $J_\mu$ with respect
to its momentum $q$. The transverse component $J_\mu^\perp$ is then
defined through the relation $q^\mu J_\mu^\perp=0$. In a similar vein in
HQET one chooses a time-like four-velocity $v_\mu$  with $v^2=1$ which
serves to define longitudinal and transverse components of a general
four-vector $a_\mu$ according to $a_\mu^\parallel=a\cdot v v_\mu$ and
$a_\mu^\perp=a_\mu - a\cdot v v_\mu$. In the dynamical formulation of HQET
this splitting up of four-dimensional space-time becomes an important
dynamical concept in as much as the covariant derivative $D_\mu$ is
split up into its longitudinal piece $D_\mu^\parallel=D\cdot v v_\mu$
and its transverse piece $D_\mu^\perp=D_\mu-D\cdot v v_\mu$. In the $v$
rest frame $v_\mu=(1,0,0,0)$ the longitudinal and transvere covariant
derivatives reduce to the time-derivative $D_0$ and the three-derivative
$\vec{D}$, respectively. The $1/m_Q$ expansion of the theory must then
be organized in such a way that in the rest frame of $v$ higher order
derivatives involve only the three-derivative $\vec{D}$ whereas the
time-derivative $D_0$ must appear only to first order \cite{HQET}.

In the two examples discussed before (two- and three-point function)
orthogonal space was $(d-1)$- and $(d-2)$-dimensional. The continuation
of space-time away from $d=4$ facilitates the regularization of
ultraviolet (UV) and IR/M singularities. All one-loop calculations in
\XLOOPS\ are in
fact carried out in $d\neq 4$ dimensions in order to be able to
control the UV and IR/M divergences. As it turns out the two-loop
integrations cannot be done in $d\neq 4$ dimensions in the generality
that is needed for the implementation in \XLOOPS. Instead the divergent
pieces of the integrands are first identified. One then subtracts and adds
auxiliary integrand functions that have the same degree of divergence
but which are sufficiently simple to allow for their integration in
$d\neq 4$ dimensions. The divergent behaviour of the original
integrand is thereby eradicated such that its numerical integration
can be performed in $d=4$ dimensions. The divergent auxiliary function
is then integrated using dimensional regularisation to the required accuracy
in powers of $\varepsilon=(d-4)/2$.

In the UV case the auxiliary integrand functions are constructed from
the same Feynman diagram which the auxiliary integrand functions are
sought to reproduce in their UV behaviour,
however, with some of the momenta and masses in the diagram
set to zero. It is clear that one has to be careful to avoid introducing
spurious IR/M divergences in the course of this procedure. The
construction of the requisite auxiliary integrand functions has been
fully automated and is implemented in \XLOOPS.

If a diagram contains IR/M divergences one
proceeds in a similar fashion. However, the construction of the
requisite auxiliary integrand functions is not as simple as in the
UV case. As of yet we have not been able to solve the
problem of constructing the auxiliary integrand functions in all generality.
There is a certain amount of progress, though,
in that we have been able to demonstrate the feasibility of this approach
for some specific and important cases. The two cases treated so far are the
two-loop $O(\alpha_s^2)$ contributions to the $b \rightarrow c$ transition
at zero recoil and the two-loop contributions to $Z \rightarrow b \bar{b}$.

There are altogether 13 two-loop contributions to the transition $b
\rightarrow c$. According to the Lee-Nauenberg theorem the sum of the
13 two-loop contributions are IR finite at zero recoil because the
accompanying Bremsstrahlung diagrams are kinematically suppressed at
this point and can therefore not be called upon for IR cancellation. But
taken separately single diagrams generally do have IR singularities
which have to be regularized. The requisite set of auxiliary integrand
functions have been explicitly constructed. After subtraction the
remaining IR finite contributions were integrated analytically.
We mention that the corresponding one-loop calculation had been done
in the early 80's for any value of the momentum transfer
variable \cite{PaschalisGounaris}. The first complete analytical zero
recoil two-loop calculation was published this year by Czarnecki and
Melnikov\cite{Czarnecki,CM}.

\begin{figure}[htbp]
  \begin{center}
    \epsfig{file=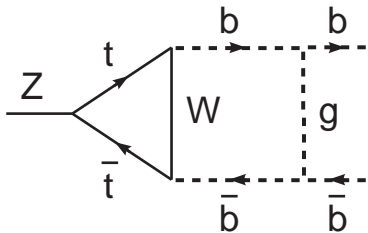,width=0.3\linewidth}
  \end{center}
 \caption[]{Parallel ladder two-loop diagram contributing to
            $Z \to b \bar{b}$.}
 \label{fig:zbb}
\end{figure}

The second example concerns the determination of the $O(g^2 \alpha_s)$
mixed electroweak and QCD two-loop contributions to
$Z \rightarrow b \bar{b}$. These radiative corrections are important for
high precision tests of the Standard Model. In Fig.\ref{fig:zbb} we
show one of the contributing diagrams which is of ``parallel ladder"
type. In our first exploratory calculation we neglected spin effects
and calculated only the scalar parallel ladder diagram. The diagram is
IR divergent, and, for $m_b=0$, also M divergent. The IR/M behaviour of
this diagram and a few other parallel ladder diagrams was improved in the
manner described above. The remaining IR/M finite integration was computed
numerically using the two-fold integral representation of massive two-loop
contributions which is also available in \XLOOPS. The auxiliary integrand
functions were then integrated in $d \neq 4$ dimensions obtaining its
IR/M divergent piece as well as its finite contribution. In order to
check on the results of this numerical evaluation the contribution of the
same diagram was then calculated using completely different integration
methods, namely the improved small momentum expansion method developed in
Refs.\cite{Bielefeld}. The numerical results were compared and satisfactory
agreement was found \cite{FFKKSST}.

For the remaining part of this presentation we presume that the
UV and IR/M improvement program for the integrands has been carried
through such that there is
no need for regularization.
The remaining discussion will be centered around integration methods
in $d=4$ dimensions.

\section{Tensor integrals and strings of {\boldmath$\gamma$}-matrices}

\begin{figure}[htbp]
  \centerline{\epsffile{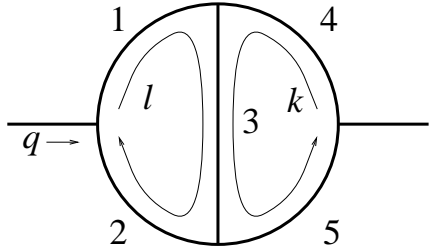}}
 \caption[]{Two-loop two-point function: the master topology}
 \label{fig:newmaster}
\end{figure}

In order to illustrate the advantages of using parallel/orthogonal
space techniques we return to the two-loop two-point case.
In Fig.\ref{fig:newmaster}
we have drawn the generic diagram of the so called master topology
which is the most difficult diagram in the
two-loop two-point category. Included are also the flow of the
loop momenta $l$ and $k$. The propagator line $3$ contains both
loop momenta $l$ and $k$ and will therefore be referred to as mixed
propagator line. The other propagator lines $1,2$ and $4,5$ contain only
one loop momentum, namely $l$ and $k$, respectively. Assuming that
$q$ is time-like, the diagram
is evaluated in the $q$-rest frame where $q^\mu=(q^0,0,0,0)$.
(If $q$ is space-like the corresponding contribution can be obtained
by analytic continuation from the time-like region.)
The loop momenta split up into their parallel and orthogonal space
components according to $l^\mu=(l^0,\vec{l}_\perp)$ and
$k^\mu=(k^0,\vec{k}_\perp)$.

Assume now that there is a $\gamma$-matrix string associated with the
diagram. Remembering that parallel and orthogonal $\gamma$-matrices
mutually anticommute and that they anticommute with $\gamma_5$ the string
can always be arranged into pieces of the form
\begin{equation}
\label{dirac}
\gamma_5^{n_5} \gamma_0^{n_0} \vec{\gamma}^\perp
\ldots\vec{\gamma}^\perp
\end{equation}
where the $n_5$ $\gamma_5$'s have been moved to the left
end of the string and the $n_0$ $\gamma_0$'s  to second to
left. The products of $\gamma_5$'s and $\gamma_0$'s can then be
simplified by using $\gamma_5^2=1$ and $\gamma_0^2=1$. If one
has an even string of $\vec{\gamma}^\perp$-matrices the string of
$\vec{\gamma}^\perp$-matrices can be replaced by products of orthogonal
space metric tensors whose weights are determined by the trace of the
$\vec{\gamma}^\perp$-matrices. For those $\vec{\gamma}^\perp$-matrices
that are contracted with the orthogonal space loop momenta $\vec{l}_\perp$
and $\vec{k}_\perp$ an appropriate symmetrization has to be carried out
w.r.t. multiply occurring loop momentum indices.  That is, returning to
the covariant description, one has the replacement
\begin{equation}
\gamma_{\mu_1}^\perp \ldots \gamma_{\mu_n}^\perp
\Rightarrow g_{{\mu_1}{\mu_2}}^\perp \ldots g_{{\mu_{n-1}}{\mu_n}}^\perp
+\ldots+ \mbox{symmetrization}
\end{equation}
Odd strings of $\vec{\gamma}^\perp$-matrices can be seen to give zero
results after integration. The case that the two-loop two-point master
diagram carries external tensor indices will not be discussed further
here but similar simplifications on the $\gamma$-matrix algebra can be
seen to occur.

Returning to Eq.(\ref{dirac}) the Dirac-string object $\sll\slk$ can be
replaced by $l_0 k_0- \vec{l}_\perp \cdot \vec{k}_\perp$. The orthogonal space
scalar product $\vec{l}_\perp \cdot \vec{k}_\perp$ contains the cosine
of the polar angle between $\vec{l}_\perp$ and $\vec{k}_\perp$. For
further processing it is advantageous to get rid of the polar
angle dependence through cancellation with the mixed propagator
$P_3=(l+k)^2-m_3^2$. This can be achieved by writing

\begin{equation}
\vec{l}_\perp \cdot \vec{k}_\perp=\Frac{1}{2}(l_0^2+k_0^2
-\vec{l}_\perp^2-\vec{k}_\perp^2-m_3^2+2l_0k_0-P_3)
\end{equation}
The mixed propagator factor $P_3$ cancels against the corresponding
denominator factor and one remains with a factorizable two-loop
contribution. The other terms are genuine two-loop contributions. However,
they no longer possess a $\cos\theta$ dependence in the integrand
numerator. They can therefore be further processed as in the scalar case
to be discussed in the next section.

If there are external indices attached to a diagram one has to solve the
problem of how the loop contributions distribute themselves onto
the various covariants that determine the covariance structure
of the diagram. Again this problem is solved very elegantly using
parallel/orthogonal space techniques. As an illustration we choose the
simple example of the two-point function carrying two external
Lorentz indices corresponding to the self-energy of an off-shell boson.
The covariance structure of the resulting contribution has the form
$C_1g_{\mu\nu}+ C_2 q_\mu q_\nu$. The invariant $C_1$ can be picked out
by calculating the orthogonal component of the two-loop integral since
$q_\mu q_\nu$ has no orthogonal space component. After
having calculated $C_1$ the invariant $C_2$ can then be obtained by considering
the parallel space component of the two-loop integral. In
the language of contractions the
invariant $C_1$ is obtained by contraction with $g^{\mu\nu}_\perp$ since
$q_\mu q_\nu g^{\mu\nu}_\perp =0$. Once $C_1$ is known the
invariant $C_2$ can
be obtained by contraction with the parallel space tensor
$q_\mu q_\nu$. Admittedly the example that we have discussed is of
a rather simple nature but it still serves to illustrate the
important point that the parallel/orthogonal space method allows one to
efficiently evaluate tensor loop integrals. Much of the tedious
algebra of matrix inversions needed in the Passarino-Veltman method
of calculating tensor integrals is avoided since the use of
parallel/orthogonal space techniques lead to an almost complete
diagonalization of the problem from the outset.

\section{The two-loop two-point function:\\ the master diagram}

Here we give a brief review of the derivation of Kreimer's two-dimensional
integral representation \cite{Kreimer2L2P} for the scalar master diagram
(see Fig.~\ref{fig:newmaster}). This is mainly intended to be a warming-up
exercise before we discuss the non-planar three-point function in the
next section.

The master diagram depends on one external
momentum, $q$, and five masses $m_i$, $(i=1,\ldots,5)$. In the scalar
case, it is UV-finite, and can therefore be calculated in $d=4$ without
any subtractions:
\begin{equation}
\label{eq:masterdef}
I \; = \; I(q^2;m_1,\ldots,m_5) \;  =
 \; \int \int \mbox{d}^4 k  \; \mbox{d}^4 l
 \; \frac{1}{P_1 P_2 P_3 P_4 P_5} \; ,
\end{equation}
where
\begin{eqnarray}
P_1 & = & {l}^2     - m_1^2 + i \rho \nonumber \\
P_2 & = & {(l-q)}^2 - m_2^2 + i \rho \nonumber \\
P_3 & = & {(k+l)}^2 - m_3^2 + i \rho \nonumber \\
P_4 & = & {k}^2     - m_4^2 + i \rho \nonumber \\
P_5 & = & {(k+q)}^2 - m_5^2 + i \rho \; .
\end{eqnarray}
For the time being we shall assume that the external momentum is time-like
$(q^2>0)$, so that we can work in its rest frame, in which
$q = (q_0,\vec{0})$,
$k = (k_0,\vec{k}_\perp)$ and
$l = (l_0,\vec{l}_\perp)$.
Introducing $k_\perp=|\vec{k}_\perp|$, $l_\perp=|\vec{l}_\perp|$ and
the polar angle $\theta$ between $\vec{k}_\perp$ and $\vec{l}_\perp$, we
rewrite the propagators as
\begin{eqnarray}
P_1 = \omega_1^2 - l_\perp^2 ,
&& \omega_1 = \sqrt{l_0^2 - m_1^2 + i \rho} \nonumber \\
P_2 = \omega_2^2 - l_\perp^2 ,
&& \omega_2 = \sqrt{(l_0-q_0)^2 - m_2^2 + i \rho} \nonumber \\
P_3 = \omega_3^2 - k_\perp^2 - l_\perp^2 - 2 k_\perp l_\perp \cos \theta ,
&& \omega_3 = \sqrt{(k_0+l_0)^2 - m_3^2 + i \rho} \nonumber \\
P_4 = \omega_4^2 - k_\perp^2 ,
&& \omega_4 = \sqrt{k_0^2 - m_4^2 + i \rho} \nonumber \\
P_5 = \omega_5^2 - k_\perp^2 ,
&& \omega_5 = \sqrt{(k_0+q_0)^2 - m_5^2 + i \rho} .
\end{eqnarray}
The important points to note here are that only the mixed propagator,
$P_3$, depends on $\theta$, and that the differences $P_1-P_2$
and $P_4-P_5$ only depend on the parallel space variables
$k_0$ and $l_0$. After performing three trivial angular
integrations, the integral (\ref{eq:masterdef}) becomes
\begin{eqnarray}
\label{eq:masterPF}
I & = & 8 \pi^2
 \int_{-\infty}^{\infty} \mbox{d} k_0
 \int_{-\infty}^{\infty} \mbox{d} l_0 \;
\frac{1}{(P_1-P_2)(P_4-P_5)}
 \int_0^{\infty} k_\perp^2 \, \mbox{d} k_\perp
 \int_0^{\infty} l_\perp^2 \, \mbox{d} l_\perp
\nonumber \\
& \times &
 \int_{-1}^{1} \mbox{d} \cos \theta
\left( \frac{1}{P_1 P_3 P_4} - \frac{1}{P_1 P_3 P_5}
     + \frac{1}{P_2 P_3 P_5} - \frac{1}{P_2 P_3 P_4}
\right).
\end{eqnarray}

Let us concentrate on the first term\footnote{
We are cheating slightly because the integral of
$\frac{1}{P_1 P_3 P_4}$ by itself diverges. However, the
divergence cancels if the four terms in (\ref{eq:masterPF})
are taken together.}
in the parantheses on the second line of Eq.~(\ref{eq:masterPF}).
The integral over the orthogonal
space variables can be regarded as a vacuum integral with ``masses''
$\omega_1$, $\omega_3$, $\omega_4$ in $d=3$.  The $z$ integral is
elementary:
\begin{eqnarray}
&&
 \int_0^{\infty} k_\perp^2 \, \mbox{d} k_\perp
 \int_0^{\infty} l_\perp^2 \, \mbox{d} l_\perp
\frac{1}{P_1  P_4}
  \int_{-1}^{1} \mbox{d} \cos \theta \frac{1}{P_3}
\nonumber \\
&=&
 \int_0^{\infty} k_\perp^2 \, \mbox{d} k_\perp
 \int_0^{\infty} l_\perp^2 \, \mbox{d} l_\perp
\frac{1}{P_1  P_4}
\frac{1}{2 k_\perp l_\perp}
\log \left[ \frac{(l_\perp-k_\perp)^2 - \omega_3^2}
                 {(l_\perp+k_\perp)^2 - \omega_3^2} \right]
\nonumber \\
&=&
 - \frac{1}{2}
 \int_{-\infty}^{\infty} \mbox{d} k_\perp
 \int_{-\infty}^{\infty} \mbox{d} l_\perp
 \frac{k_\perp l_\perp}
      {(l_\perp^2 - \omega_1^2)(k_\perp^2 - \omega_4^2)}
 \log \left[ l_\perp + k_\perp + \omega_3 \right] \, .
\end{eqnarray}
Now, both the $k_\perp$ and
$l_\perp$ integrations can be performed by closing the integration contours
in the complex plane (taking care not to
cross the cut of the logarithm) and applying Cauchy's theorem,
which gives:
\begin{eqnarray}
&&
 - \frac{(2 \pi i)^2}{8}
 \log \left[ \omega_1 + \omega_3 + \omega_4 \right] .
\end{eqnarray}

Collecting the four terms in (\ref{eq:masterPF}) and expressing
everything in terms of the dimensionless variables $x=l_0/q_0$, $y=k_0/q_0$,
we finally obtain:
\begin{equation}
\label{eq:masterfinal}
I = \frac{4 \pi^4}{q^2}
 \int_{-\infty}^{\infty} \mbox{d} x
 \int_{-\infty}^{\infty} \mbox{d} y \;
 \frac{1}{w_1^2-w_2^2} \frac{1}{w_4^2-w_5^2}
 \log \left[ \frac{(w_1+w_3+w_4)(w_2+w_3+w_5)}
                  {(w_2+w_3+w_4)(w_1+w_3+w_5)}
      \right]
\end{equation}
with
\begin{eqnarray}
w_1 = \sqrt{x^2     - \frac{m_1^2}{q^2} +i \rho}
&&
w_2 = \sqrt{(x-1)^2 - \frac{m_2^2}{q^2} +i \rho}
\nonumber \\
w_3 = \sqrt{(x+y)^2 - \frac{m_3^2}{q^2} +i \rho}
&&
w_4 = \sqrt{y^2     - \frac{m_4^2}{q^2} +i \rho}
\nonumber \\
w_5 = \sqrt{(y+1)^2 - \frac{m_5^2}{q^2} +i \rho} .
&&
\end{eqnarray}

The representation (\ref{eq:masterfinal}) is valid for arbitrary masses,
and it gives the correct analytic continuation to space-like
external momenta. Except for a few very special threshold configurations,
it is possible to take the limit $\rho \to 0$ under the integral sign,
yielding an integrand which is continuous in the entire $(x,y)$ plane
and can easily be integrated numerically. An attractive feature of this
result is that we do not have to distinguish whether $q^2$ is below,
in between, or above the thresholds: one simple formula covers all
cases. Eq.(\ref{eq:masterfinal}) also gives the correct imaginary part.

Tensor integrals (of rank greater than one) require subtractions to
make them UV-convergent. \XLOOPS\ automatically performs the necessary
subtractions \cite{Kreimer2LTensors} and then generates two-dimensional
numerical integrands for the finite remainders that are very similar to
the one derived here for the scalar case. The only difference is that
they contain additional factors of $x$, $y$, or $w_i$ in the numerator
which are not difficult to deal with.

\section{The two-loop three-point function:\\ the crossed
ladder configuration}

\begin{figure}[htbp]
  \centerline{\epsffile{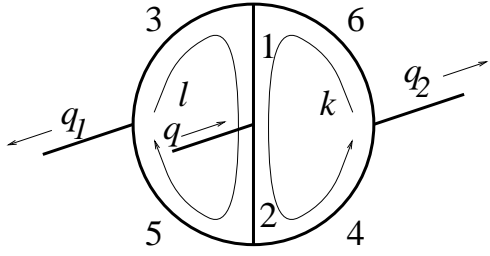}}
 \caption[]{The two-loop crossed vertex diagram}
 \label{fig:crossedvertex}
\end{figure}

\begin{figure}[htbp]
  \begin{center}
    \epsfig{file=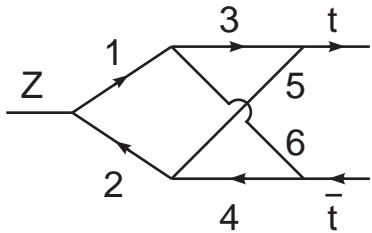,width=0.3\linewidth}
  \end{center}
 \caption[]{Crossed ladder two-loop contribution to
              $Z(\mbox{off-shell}) \rightarrow t \bar{t}$.}
 \label{fig:japan}
\end{figure}

Starting from the two-point master topology, one obtains the crossed
ladder topology by attaching the third leg to the central propagator,
as shown in Fig.\ref{fig:crossedvertex}. This is the most complicated
two-loop three-point topology because, no matter how the loop momenta $k$
and $l$ are routed, one cannot avoid having {\em two} mixed propagators
that depend on the scalar product $k\cdot l$. Another way of drawing
this topology is shown in Fig.\ref{fig:japan}.

In this section, we will give a brief and somewhat simplified account
of how a two-fold integral representation for this diagram can be
obtained. For a more detailed explanation, and for other two-loop
three-point topologies, we refer to \cite{FKK,XLoops}.

The scalar integral is defined by
\begin{equation}
V \; = \;
V(q_1^2,q_2^2,q^2;m_1,\ldots,m_6) \;  =
 \; \int \int \mbox{d}^4 k  \; \mbox{d}^4 l
 \; \frac{1}{P_1 P_2 P_3 P_4 P_5 P_6} \; ,
\end{equation}
where $q=q_1+q_2$ and the propagators are
\begin{eqnarray}
P_1 & = & {(k+l-q_1)}^2 - m_1^2 + i \rho \nonumber \\
P_2 & = & {(k+l+q_2)}^2 - m_2^2 + i \rho \nonumber \\
P_3 & = & {(l-q_1)}^2   - m_3^2 + i \rho \nonumber \\
P_4 & = & {(k+q_2)}^2   - m_4^2 + i \rho \nonumber \\
P_5 & = & {l}^2         - m_5^2 + i \rho \nonumber \\
P_6 & = & {k}^2         - m_6^2 + i \rho \; .
\end{eqnarray}

\begin{figure}[htbp]
  \begin{center}
    \epsfig{file=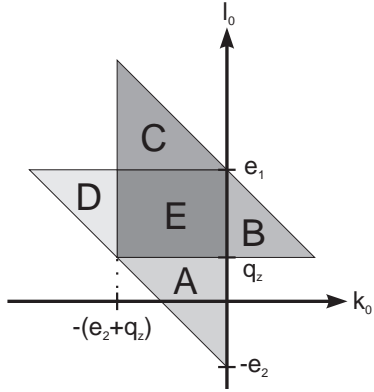,width=0.3\linewidth}
  \end{center}
 \caption[]{Integration region in the $(\tilde{k}_0,\tilde{l}_0)$ plane}
 \label{fig:k0l0region}
\end{figure}

We use a frame where $q = (e_1+e_2, 0,\vec{0})$,
$q_1 = (e_1, q_z,\vec{0})$,
$q_2 = (e_2,-q_z,\vec{0})$,
$k = (k_0,k_1,\vec{k}_\perp)$ and $l = (l_0,l_1,\vec{l}_\perp)$.
This time, $\vec{k}_\perp$ and $\vec{l}_\perp$ are two-component
vectors.
Introducing polar coordinates in the orthogonal space and carrying
out one trivial angular integration, we find that
\begin{eqnarray}
 &&   \int \mbox{d}^4 k \; \int \mbox{d}^4 l \ldots
\nonumber \\
& = & 2 \pi
 \int_{-\infty}^{\infty} \mbox{d} k_0
 \int_{-\infty}^{\infty} \mbox{d} l_0 \;
 \int_{-\infty}^{\infty} \mbox{d} k_1 \;
 \int_{-\infty}^{\infty} \mbox{d} l_1 \;
 \int_0^{\infty} k_\perp \, \mbox{d} k_\perp
 \int_0^{\infty} l_\perp \, \mbox{d} l_\perp
 \int_0^{2\pi} \mbox{d} \theta \ldots
\nonumber \\
& = & \frac{\pi}{2}
 \int_{-\infty}^{\infty} \mbox{d} \tilde{k}_0
 \int_{-\infty}^{\infty} \mbox{d} \tilde{l}_0 \;
 \int_0^{\infty} \mbox{d} s
 \int_0^{\infty} \mbox{d} t
 \int_{-\infty}^{\infty} \mbox{d} k_1 \;
 \int_{-\infty}^{\infty} \mbox{d} l_1 \;
 \int_0^{2\pi} \mbox{d} \theta \ldots \, ,
\end{eqnarray}
where $s=l_\perp^2$ and $t=k_\perp^2$. In the last line, we have replaced
$k_0$ and $l_0$ with the shifted variables $\tilde{k}_0=k_0-k_1$ and
$\tilde{l}_0=l_0-l_1$. As a result, all the propagators become linear
in $k_1$ and $l_1$, e.g.
\begin{equation}
\label{eq:p6tilde}
P_6 = k_0^2 - k_1^2 - t - m_6^2 + i \rho
  = \tilde{k}_0^2 + 2 \tilde{k}_0 k_1 - t - m_6^2 + i \rho \, .
\end{equation}
Thus, the $k_1$ and $l_1$ integrations can be performed
by closing the contours in either the upper or lower half
plane and applying Cauchy's theorem. In order to understand
the general structure of the result of using Cauchy's theorem, let us decompose
the integral into $(2\times 2\times 2=8)$ partial fractions of the form
\begin{equation}
 \int_{-\infty}^{\infty} \mbox{d} k_1 \;
 \int_{-\infty}^{\infty} \mbox{d} l_1 \;
 \frac{1}{(k_1-z_1)(l_1-z_2)(k_1+l_1+z_3)}
= \frac{(2\pi i)^2}{z_1+z_2+z_3} .
\end{equation}
This equation holds when the signs of the imaginary parts of
$z_1$, $z_2$ and $z_3$ are all equal. Otherwise, the integral
on the left hand side vanishes. Now the imaginary parts of the
$z_i$'s all come from the infinitesimal $i \rho$, divided by
linear combinations of $\tilde{k}_0$, $\tilde{l}_0$ and the
external momenta (cf. eg. eq.~(\ref{eq:p6tilde})). As a consequence,
each term of the partial fraction decomposition only gives
a non-vanishing contribution inside a finite, triangular
region in the $(\tilde{k}_0,\tilde{l}_0)$ plane. The union
of those triangles is shown in Fig.~\ref{fig:k0l0region}.

After a further partial fraction decomposition wrt.\ $\cos \theta$,
the $\theta$-integration can be carried out using
\begin{equation}
 \int_0^{2\pi} \mbox{d} \theta
 \frac{1}{A - B \cos \theta}
 = \frac{2\pi}{\sqrt{A^2-B^2}}
\end{equation}
(and being careful to pick the correct branch of the square root
on the right hand side!).

Having performed the $k_1$, $l_1$ and $\theta$ integrations\footnote{
For technical reasons, the $\theta$ integration in \cite{FKK} is done first.
The result is, of course, the same.},
one obtains the following intermediate expression:
\begin{equation}
\label{eq:Vintermediate}
 V \; = \; \sum_j
\frac{ C_j \theta_j }
     { \sqrt{(a_j t + b_j + c_j s)^2 - 4 st} }
\prod_{i=1}^3 \frac{1}
         {(\tilde{a}_{ij} t + \tilde{b}_{ij} + \tilde{c}_{ij} s)} \, .
\end{equation}
The quantities $C_j$, $a_j$, $b_j$, $c_j$, $\tilde{a}_{ij}$,
$\tilde{b}_{ij}$ and $\tilde{c}_{ij}$ are all rational functions
of $\tilde{k}_0$, $\tilde{l}_0$, $q_i$ and $m_i$. $\theta_j$ is
a product of step functions restricting $\tilde{k}_0$ and
$\tilde{l}_0$ to one of the triangles discussed above.

By means of Euler's change of variables, it is possible to
get rid of the square root in (\ref{eq:Vintermediate}) and
perform both the $s$ and $t$ integrals analytically, leading,
eventually, to dilogarithms and Clausen functions. Although the final
expression is too long to be listed here in explicit form, it is a
sufficiently well-behaved function to allow the remaining $\tilde{k}_0$ and
$\tilde{l}_0$ integrations to be evaluated numerically without problems.

\section{Checks on two-loop three-point functions}

We have performed a number of checks on the integration methods that
go into \XLOOPS\ as well as on the performance of \XLOOPS\ itself. Some of
these checks have been discussed before, as e.g. checks on the
two-loop $O(\alpha_s^2)$ contributions to the $b \rightarrow c$ transition
at zero recoil.

As concerns the two-loop three-point function the checks divide up into
checks on the correctness of the theoretical structure of the results and
numerical cross-checks against the results of other authors for specific
mass cases. For example, we have checked that the two-fold integral
representations of the parallel and crossed ladder contributions have the

\begin{itemize}
\item correct cut structure in the analytic plane as prescribed by
the Landau-Cutkosky cutting rules
\item possess the correct anomalous threshold behaviour at the
prescribed anomalous thresholds
\end{itemize}

We have looked at all two-particle cuts and, with particular care, at
all three-particle cuts and have found that the analyticity structure
of our parallel and crossed ladder two-loop results are in complete
agreement with the Landau-Cutkosky cutting rules.

As a check on the correct anomalous threshold behaviour we ran our
numerical program across an anomalous threshold. We start with the
one-loop vertex contribution which possesses anomalous thresholds at
\begin{equation}
1+2\mu_1\mu_2\mu_3-\mu_1^2-\mu_2^2-\mu_3^2=0
\end{equation}
where ($i=1,2,3$)
\begin{equation}
\mu_i=\frac{(m_1^2+m_2^2+m_3^2-m_i^2-p_i^2)m_i}{2m_1m_2m_3}
\end{equation}
As sample values we took the following set of masses and momenta
(in arbitrary units)
\begin{equation}
m_1=m_2=m_3=1
\end{equation}
\begin{equation}
p_1^2=6,\, p_2^2=5,\, p_3^2=-4-\sqrt{15}\approx-7.873
\end{equation}

\begin{figure}[htb]
\[
\begin{array}{cc}
\epsfig{file=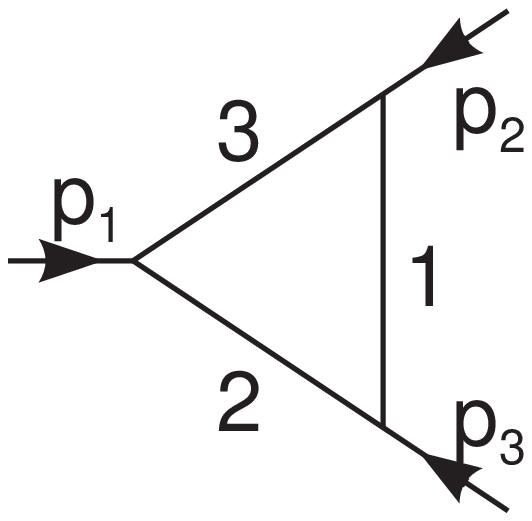,width=0.18\linewidth} &
\epsfig{file=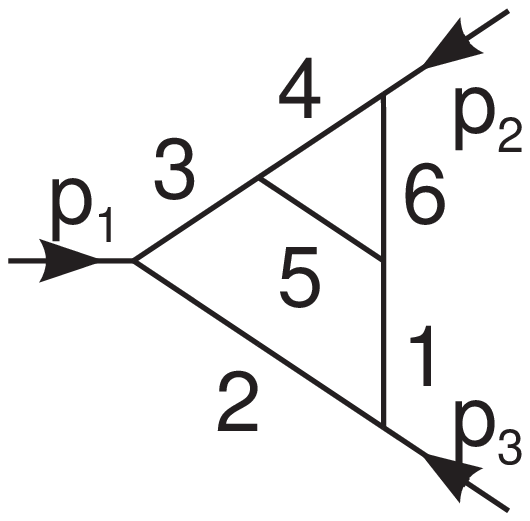,width=0.18\linewidth} \\
\epsfig{file=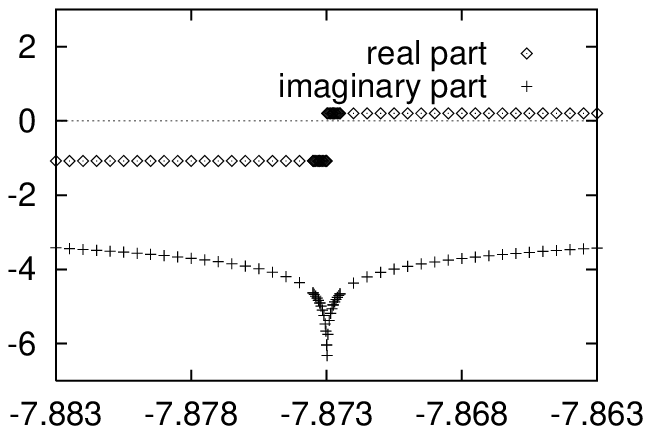,width=0.4\linewidth} &
\epsfig{file=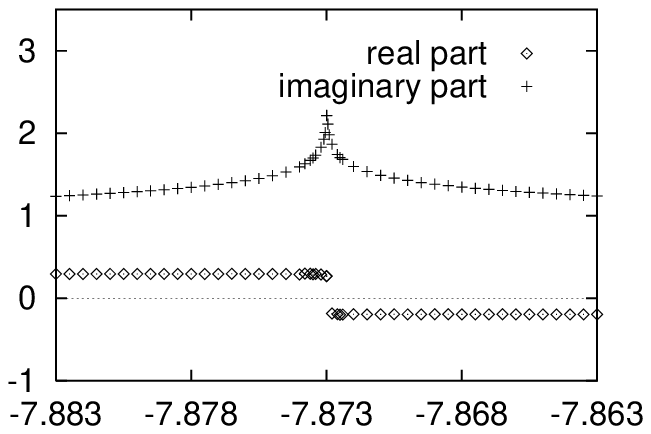,width=0.4\linewidth} \\
{\scriptstyle{\sf p_3^2}} &
{\scriptstyle{\sf p_3^2}}
\end{array}
\]
\caption{Plot of real and imaginary parts of a vertex function in the
vicinity of an anomalous threshold, one-loop (left) and two-loop
(right)}
\label{fig:anom}
\end{figure}

In Fig.\ref{fig:anom} we show the values of the imaginary and the real
part\footnote{Because our vertex function contains an extra factor of $i$
the role of the imaginary and real part is interchanged relative to what
one is used to}of the vertex contribution in the vicinity of the anomalous
threshold at $p_3^2\approx-7.873$. The plot clearly shows the
logarithmic type singularity of an anomalous threshold at $x=x_0$
which reads
\begin{equation}
i \ln(x-x_0) + C(x) = i \ln |x-x_0| \, \pm \, \pi\theta(x-x_0) + C(x)
\end{equation}
where $C(x)$ is a complex-valued function that is regular in the vicinity
of $x=x_0$.
We also checked on the anomalous threshold behaviour of the corresponding
parallel ladder two-loop vertex correction. The two-loop vertex correction
has an anomalous threshold at the same position $p_3^2\approx-7.873$
for any values of the masses $m_4,m_5,m_6$. The numerical evaluation
of the two-loop graph showed the same behaviour around the anomalous
threshold as the one-loop graph (see Fig.\ref{fig:anom}). The numerical
stability in the direct vicinity of the anomalous threshold was found to
be satisfactory in both cases but was somewhat decreased in the two-loop
case relative to the one-loop case.

Further we ran our program for specific mass configurations and compared
the numerical results with the results of other groups.  For example,
we evaluated the crossed ladder vertex graph for external momenta
with $q^2=p_1^2=p_2^2 \neq 0$ and let the internal masses $m_i=m (i=1,
\ldots,6)$ go to zero. When $m=0$ Ussyukina and Davydychev have shown
that the crossed ladder configuration can be calculated analytically,
and that the result turns out to be equal to the square of a one-loop
graph \cite{UD}. Our numerical $m \rightarrow 0$ results agree with
the Ussyukina-Davydychev limit.  It came as a pleasant surprise that
numerical stability was retained in this limit.

Next we took the specific momentum configuration $p_1^2=p_2^2=0$ and
set all six internal masses in the crossed ladder configuration to
$m=150$\,GeV.  We compared our results with the corresponding calculation
by Tarasov \cite{Tarasov}, who used an improved small momentum expansion
using conformal mapping and Pad\'e approximants. There was full agreement
on the real and imaginary parts of the crossed ladder configuration
above, below and close to threshold $q^2=4m^2$. In particular there was
no noticeable decline in numerical stability close to threshold which had
been a troublesome region in the low momentum expansion method before it
was improved by conformal mapping. It is not so simple to compare the time
performance of the two approaches since the low momentum expansion method
requires a great amount of preparatory work in that the coefficients of
the low momentum expansion have to be calculated first.

We also compared our results to an evaluation of the crossed ladder
contribution to $Z(\mbox{off-shell}) \rightarrow t \bar{t}$ done by
the Japanese KEK group Fujimoto et al.\cite{Fujimoto}. The relevant mass
configuration
is $m_1=m_2=m_3=m_4=\sqrt{q_1^2}=\sqrt{q_2^2}=150$\,GeV (the would-be
mass of the top quark in 1994) and $m_5=m_6=91$\,GeV. The contribution
clearly corresponds to the off-shell $Z t \bar{t}$ vertex with internal
$Z$ and $t$ exchange. Again we found full numerical agreement for both
the real and imaginary parts. The Japanese group used Feynman parametrization
and reduced the number of necessary numerical
loop integrations to three by analytic integration. For comparable
accuracy the necessary computing time is considerably smaller
in the Mainz approach.

As a final example of the reliability of the massive two-loop crossed
and parallel ladder integration techniques and the numerical integration
techniques based on them we discuss a recent application of these methods
to the calculation of the dominant two-loop corrections, of order ${\cal
O}(G_F^2M_H^4)$, to the partial widths of the decays of a heavy Higgs to
pairs of W- and Z-bosons, carried out in collaboration with B.A. Kniehl
and K. Riesselmann \cite{MainzMunich}. The calculation was done again
using the Equivalence Theorem, i.e. the dominant longitudinal degrees
of freedom of the gauge bosons were represented by the massless and
spinless Goldstone boson excitations of the effective theory. Of the
many contributing diagrams we show only two representative diagrams of
the class of crossed and parallel ladder configurations which are the
most difficult to calculate.

\begin{figure}[htbp]
  \begin{center}
    \epsfig{file=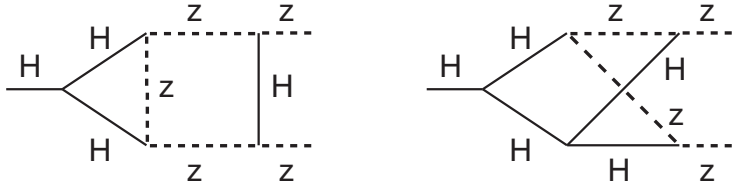,width=0.6\linewidth}
  \end{center}
 \caption[]{Two-loop parallel and crossed ladder contributions to
            $H \to ZZ$ in the effective theory. Solid lines:
            massive Higgs particles; dashed lines: massless gauge
            bosons.}
 \label{fig:hzz}
\end{figure}
In the numerical evaluation of these diagrams we started off with small
finite Goldstone masses which were then set to smaller and smaller values
and finally to zero. We found that the numerical results were stable in
this limit. After compounding the contributions of the many contributing
diagrams and carrying out the renormalization in the on-shell scheme, the
correction factor was found to be
$1+14.6\% (m_H/\mbox{TeV})^2+16.9\% (m_H/\mbox{TeV})^4$,
indicating that for $m_H>1 \mbox{TeV}$ the Standard Model ceases
to be weakly interacting. Numerically, this result agrees with that
of Ghinculov, who had used entirely different integration techniques
to evaluate the same decay process \cite{ghinculov}. From the fact that the
calculation involves a large number of massive two-loop diagrams that
contribute to the process we believe that the numerical agreement of the
two calculations provides for another strong test of the reliability of
our integration techniques and the numerical routines based on them as
well as those used by Ghinculov.
We mention that a two-loop calculation of the fermionic contributions
to $H \to ZZ,WW$ is under way.

\section{Some sample calculations with \BXLOOPS}

This summer saw the premi\`ere of a live performance of \XLOOPS\
during a talk given by J.~Franzkowski at the Klausurtagung of the Mainz
Graduiertenkolleg ``Teilchenphysik bei hohen und mittleren Energien". For
the performance he used a laptop and a projectable screen display. Nothing
can beat the excitement of a live performance of \XLOOPS\ either at a PC
session or during a talk using a projector screen.  We shall nevertheless
try to capture some of the spirit of a live demonstration of \XLOOPS\
by presenting two sample calculations done by \XLOOPS. In the first
example we calculate the on-shell one-loop self-energy of the neutral
gauge boson $Z$ resulting from the contribution of the top quark loop.
In the second example we calculate one of the mixed $O(g^2\alpha_s)$
QCD/electroweak contributions to the two-loop flavour-changing self-energy
of the $s \rightarrow d$ quarks.

\begin{figure}[htb]
  \begin{center}
    \epsfig{file=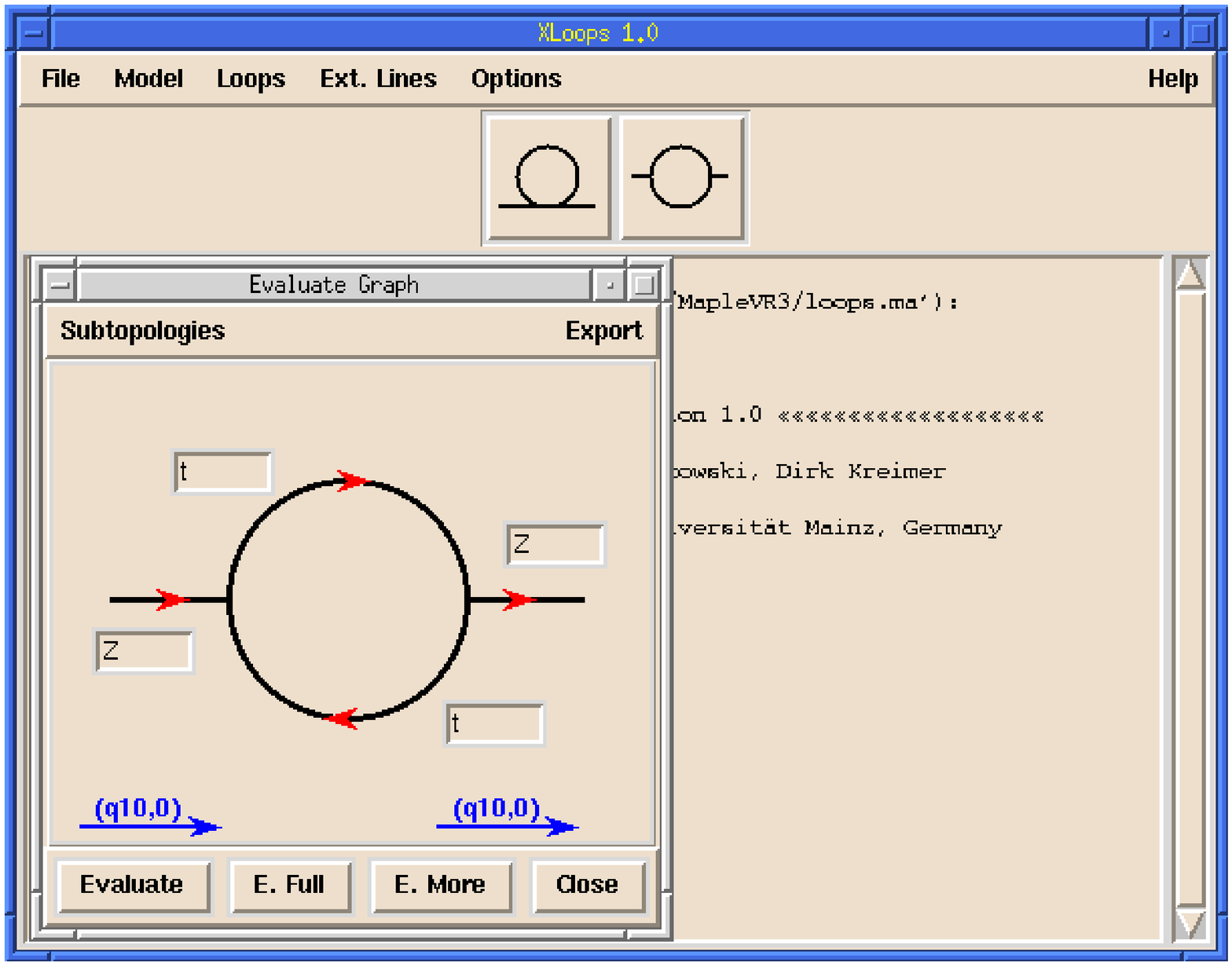,width=0.7\linewidth}
  \end{center}
  \caption[]{\XLOOPS\ display of a one-loop Feynman diagram.}
  \label{fig:bild2a}
\end{figure}

At the beginning of an \XLOOPS\ session one has to specify the model
whose Feynman rules are to be used in the loop evaluation. In both of
the examples to be discussed here one clicks on ``Standard Model" in the
``Model" menu which would call up the Feynman rules of the Standard Model.
Also the user has to specify the numbers $m$ and $n$ for his choice of
the $m$-point $n$-loop function.  In the case of the one-loop self-energy
of the $Z$ he would then click on ``two" external lines ($m=2$) and on
``one" loop ($n=1$). The topology bar would then show the two possible
two-point one-loop topologies. These can be discerned at the top of
Fig.\ref{fig:bild2a}. One then clicks on the desired topology which would
be the second of the two topologies in Fig.\ref{fig:bild2a}. After having
selected the topology the desired one-loop graph appears on the screen
in an enlarged version as shown in Fig.\ref{fig:bild2a}.

One then has to attach arrows to the lines in the graph according to
the quantum number flow (wrongly pointed arrows would lead to an error
message) and specify the particles in the graph by inserting the particle
names in the particle entry fields. In the next step one has to specify
the mass and
momentum parameters. There is an ``Insert Particle Properties" option as
a menu entry which automatically substitutes the mass values provided
by the latest version of the Particle Data Group listings. Mass values
can of course also be introduced by hand. In our example we have $m_Z=
91.187$ GeV and $m_t= 176$ GeV. The one-loop graph is evaluated in the
rest system of the on-shell $Z$. Thus one chooses $q_{10} = 91.187$ GeV.
The calculation of the graph is then executed by clicking on the
``E.(Evaluate)Full" bar. In this one-loop example the numerical answer
would appear within seconds in the following format:


{\footnotesize
\begin{verbatim}
[

C1 = [

                                                  -16
 146.22583756233107509 I, .26077710144029877423*10    - 1219.7690445131992210 I

 ],

C2 = [.00096061186286253340551 I,

                            -20
    .18797591473406467665*10    - .0080036871858407297497 I],

C1 G(nu1, nu2) + C2 q1(nu1) q1(nu2)]
\end{verbatim}
}

The numerical output is given in terms of the two invariants $C_1$ and
$C_2$ in the covariant expansion $C_1 g_{\mu \nu} + C_2 q_\mu q_\nu$. Each
of the invariants is displayed as a list $[a,b]$ which represents a series
$a/\varepsilon + b + {\cal O}(\varepsilon)$. In the present case the
coefficients $a$ are purely imaginary while the $b$'s have a tiny
real part, that would vanish if the causal $i \rho$'s in the propagators
were put to zero exactly. This corresponds to the fact that in the
present momentum configuration, below the $t \bar{t}$-threshold,
the self-energy $\Sigma$ (which differs from the diagram itself
by a factor of $i$) is purely real.

\begin{figure}[htb]
  \begin{center}
    \epsfig{file=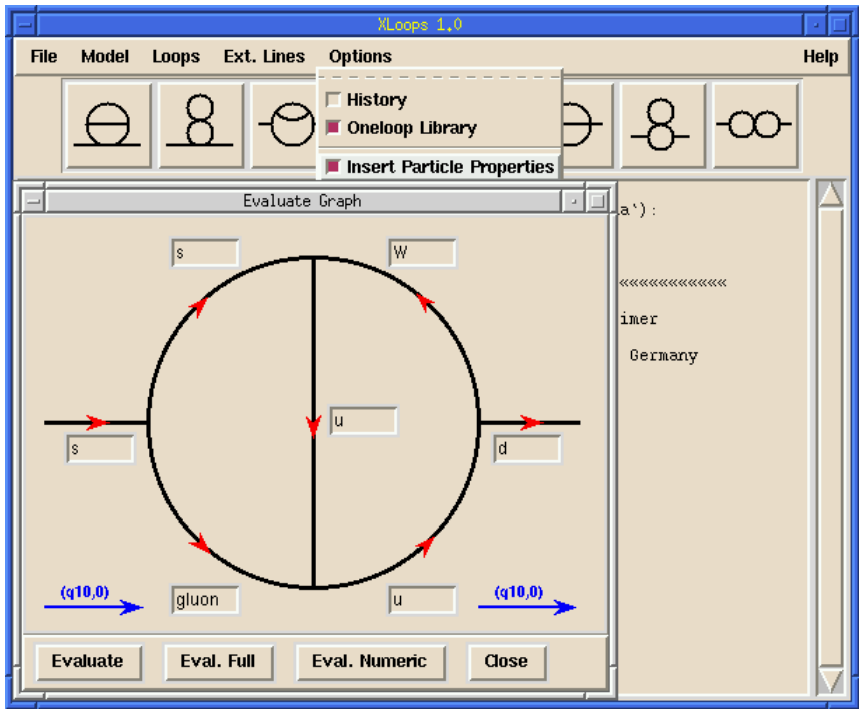,width=0.7\linewidth}
  \end{center}
  \caption[]{\XLOOPS\ input panel for the two-loop master diagram.}
  \label{fig:koern3}
\end{figure}

The second example is drawn from a recent determination of the $O(g^2
\alpha_s)$ two-loop contribution to the flavour changing self-energy
for $s \rightarrow d$. Of the contributing diagrams we take one of the
diagrams with a master topology as drawn in Fig.\ref{fig:koern3}. One
proceeds as was described in the first example only that one specifies
to ``two" in the menu ``Loops".  One recognizes some of the relevant two-loop
topologies at the top of the display depicted in Fig.\ref{fig:koern3}. In
the present example one clicks on the master topology which is the
fourth two-loop diagram from the left in the topology bar which is covered
up in the display Fig.\ref{fig:koern3}. The relevant Feynman diagram
pops up in which the quantum number flows and particle names have to be
inserted as shown in Fig.\ref{fig:koern3}. Finally, after specifying the
particle masses and the momentum of the external particle one initiates
the numerical evaluation of the diagram:

{\footnotesize
\begin{verbatim}
                                      -6
    [C1 = [ - .68103432226158921830*10   I delta3(tmu1, tmu2),

                                 -5
         .42790442626661980279*10   delta3(tmu1, tmu2)

                                        -5
              - .40233768346700174054*10   I delta3(tmu1, tmu2),

         [%1, [ - 4.98962 - 2.43436 I, .0015744 + .00190137 I]],

         .000026216132920579719619 delta3(tmu1, tmu2)

                                        -6
              + .60792896861672700256*10   I delta3(tmu1, tmu2)],

                                         -6
        C2 = [ - .68103432226158921830*10   I delta3(tmu1, tmu2),

                                    -5
            .42790442626661980279*10   delta3(tmu1, tmu2)

                                           -5
                 - .40233768346700174054*10   I delta3(tmu1, tmu2),

            [%1, [ - 4.99119 - 2.43249 I, .00176832 + .00185428 I]],

            .000026216121704383664719 delta3(tmu1, tmu2)

                                           -6
                 + .59969113686867746914*10   I delta3(tmu1, tmu2)],

        C3 = [.00015525369186016884013 I delta3(tmu1, tmu2),

             - .00051268542153584540755 delta3(tmu1, tmu2)

                 + .00072319063843461738822 I delta3(tmu1, tmu2),

            [%1, [ - 845.404 - 954.737 I, .742148 + .926344 I]],

            3.5949563175956949478 delta3(tmu1, tmu2)

                 + 96495.670995961263803 I delta3(tmu1, tmu2)],

        C4 = [.00015525369186016884013 I delta3(tmu1, tmu2),

             - .00051266860707202340491 delta3(tmu1, tmu2)

                 + .00072847640460045290512 I delta3(tmu1, tmu2),

            [%1, [ - 844.468 - 954.961 I, .576421 + .882752 I]],

            10.877778389998066059 delta3(tmu1, tmu2)

                 + 103995.51572482192447 I delta3(tmu1, tmu2)],

        C1 (1 &* ONE) + C2 (1 &* Dg5) + C3 (1 &* Dg0) + C4 &*(1, Dg5, Dg0)]

                                        -6
%1 :=           .85129290282698652282*10   delta3(tmu1, tmu2)
\end{verbatim}
}

The results are given in terms
of the four invariants $C_1$, $C_2$, $C_3$ and $C_4$ in the rest frame
decomposition
$C_1\bar{u}u
+ C_2\bar{u}\gamma_5 u
+ C_3\bar{u} \gamma_0 u
+ C_4\bar{u} \gamma_5 \gamma_0 u$.
For each of the invariants $C_j$, the data are shown as a list
$[a,b,[c,[d,e]],f]$. Their meaning is as follows:
$C_j=a/\varepsilon^2+b/\varepsilon+c\times(d\pm e)+f+{\cal O}(\varepsilon)$.
$f$ is the piece of the finite part of $C_j$ that is calculated
analytically, $d$ is the part that is obtained by numerical integration,
and $e$ is an estimate of the accuracy of the numerical integration.
The overall colour factor of the diagram \verb|delta3(tmu1, tmu2)| is
displayed separately. It is calculated from the rules of the $SU(N)$
colour algebra which are known to \XLOOPS.

\begin{figure}[htb]
  \begin{center}
    \epsfig{file=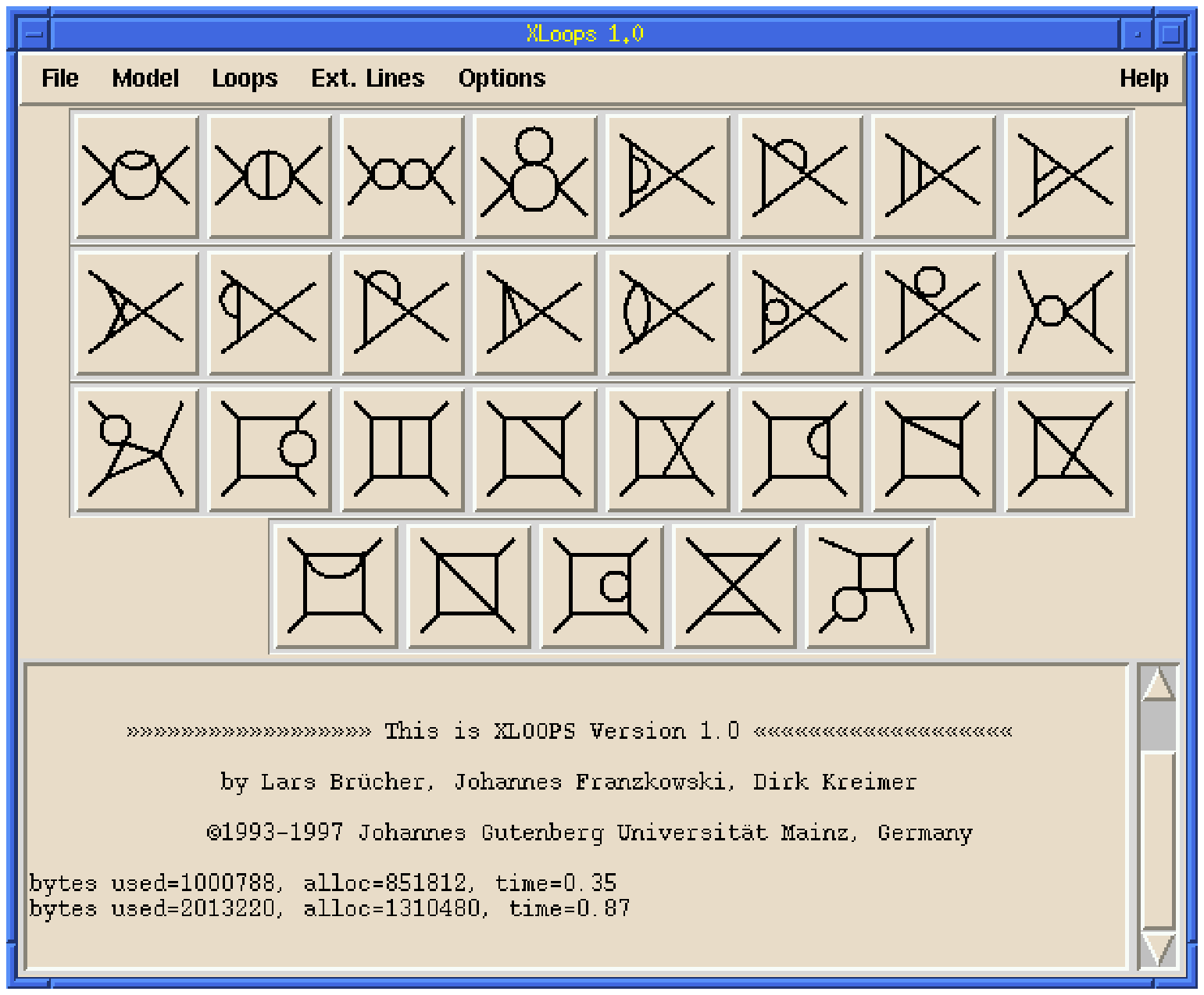,width=0.7\linewidth}
  \end{center}
  \caption[]{List of four-point two-loop topologies.}
  \label{fig:koern2}
\end{figure}

As indicated in Fig.\ref{fig:abc12} and Table 1
we have started work on massive four-point two-loop
functions. As a preparatory step we have classified all the two-loop
topologies and generated their diagrams. All the four-point two-loop
topologies are shown in Fig.\ref{fig:koern2}. We shall refer to the $p$-th
diagram in the $q$-th row by $(p,q)$. The calculation of diagram
$(2,4)$ was done first because of its implication for a two-loop
chiral perturbation theory calculation of $\gamma + \gamma \rightarrow
\pi + \pi$.
The calculation of the scalar diagram $(2,4)$ was recently completed.
The techniques of parallel and orthogonal space integrations used in
the four-point case are similar
to the three-point two-loop case only that parallel space is now
3-dimensional. Six of the
eight needed integrations
were done analytically. The remaining two-fold integral
representation was found to be suitable for numerical processing.
The calculation of diagrams $(3,3)$ and $(7,3)$ is well under way.

\section{Present status of \BXLOOPS\ and outlook}

The presently available version of \XLOOPS\ is a demonstration version
which runs
under the name of \XLOOPS\ 0.9. It is meant to whet the appetite of
the user. The demo version features all Standard Model Feynman rules
including QCD. It allows one to calculate one-loop graphs up to and
including three-point functions. It is available at the web-site
http://wwwthep.physik.uni-mainz.de/$\sim$xloops.

A full version \XLOOPS\ 1.0 has been completed and is currently being
tested. It is accompanied by a 110 page manual which can be obtained
at the above web-site. \XLOOPS\ 1.0 and the manual should be
available in a few weeks time. There are some additional
features as Feynman rules for the two Higgs doublet model. One of the
future items of \XLOOPS\ is the automatic generation of Feynman diagrams from
a given set of Feynman rules. One would then like to avail of subsets
of the Standard Model Feynman rules. This feature is already incorporated
in \XLOOPS\ 1.0 in that one can select the Feynman rules of QED and QCD.
\XLOOPS\ 1.0 incorporates first two-loop features in the form
of two-loop two-point functions including their tensor decomposition.

As time goes on more features will be added to \XLOOPS. We plan to issue
\XLOOPS\ 1.1 by the end of this year. It will contain a one-loop
four-point function routine including tensor structure capabilitiies. Also
\XLOOPS\ 1.1 will be able to automatically generate
all one- and two-loop Feynman diagrams that contribute to a given process
including a drawing routine which generates Postscript output. Also we
plan to install a plotting routine that will generate plots when input
parameter values of masses or momenta are varied continuously
over a range of the input parameter values. There are also plans to write
a longer review on the theoretical background with a detailed description
of the calculational techniques that go into \XLOOPS\ .

Work is going on on the two-loop three-point function which is
ready conceptually. More Feynman rules will be added such as those of the
minimal supersymmetric Standard Model and chiral perturbation theory.
Further in the future lies the implementation of
the two-loop four-point function which is presently under study.
Also we plan to install automatic renormalization capabilities.
The ground is prepared for renormalization in that the one-loop
routines calculate to $O(\varepsilon)$ accuracy such that
one-loop counter terms can be computed automatically. In the very far future
we hope to provide the possibility to evaluate complete processes
including phase space integration routines for tree graph contributions.
All of this needs a great deal of man power. We hope that we can count
on the support of the community in this enterprise.

\section*{Acknowledgement}

We would like to thank all members of our Mainz multi-loop group for
their collaboration and for their advice and help in preparing this
manuscript.

\end{document}